\documentclass[twocolumn,showpacs,preprintnumbers,amsmath,amssymb]{revtex4}
\usepackage{amsfonts}
\usepackage{graphicx}
\usepackage{dcolumn}
\usepackage{bm}
\input epsf
\epsfclipon

\begin{document}
\newcommand{\be}{\begin{equation}}
\newcommand{\ee}{\end{equation}}
\newcommand{\ba}{\begin{eqnarray}}
\newcommand{\ea}{\end{eqnarray}}
\newcommand{\ds}{\displaystyle}
\newcommand{\mean}[1]{\left\langle #1 \right\rangle}
\newcommand{\abs}[1]{\left| #1 \right|}
\newcommand{\la}{\langle}
\newcommand{\ra}{\rangle}
\newcommand{\RA}{\Rightarrow}
\newcommand{\tet}{\vartheta}
\newcommand{\eps}{\varepsilon}
\newcommand{\ul}[1]{\underline{#1}}
\newcommand{\non}{\nonumber \\}
\newcommand{\no}{\nonumber}
\newcommand{\eqn}[1]{Eq. (\ref{#1})}
\newcommand{\Eqn}[1]{Eq. (\ref{#1})}
\newcommand{\eqs}[2]{Eqs. (\ref{#1}), (\ref{#2})}
\newcommand{\pic}[1]{Fig. \ref{#1}}
\newcommand{\name}[1]{{\rm #1}}
\newcommand{\bib}[4]{\bibitem{#1} {\rm #2} (#4): #3.}
\newcommand{\vol}[1]{{\bf #1}}
\newcommand{\et}{{\it et al.}}
\newcommand{\fn}[1]{\footnote{ #1}}
\newcommand{\D}{\displaystyle}
\newcommand{\T}{\textstyle}
\newcommand{\lab}[1]{\label{#1}}

\title{Supremacy distribution in evolving networks}\author{Janusz A. Ho{\l}yst}
\email{jholyst@if.pw.edu.pl} \affiliation{Faculty of Physics and Center of
Excellence Complex Systems Research \\Warsaw University of Technology \\
Koszykowa 75, PL--00-662 Warsaw, Poland}
\author{Agata Fronczak}%
 \email{agatka@if.pw.edu.pl}
\affiliation{Faculty of Physics and Center of Excellence Complex Systems
Research \\Warsaw University of Technology \\ Koszykowa 75, PL--00-662 Warsaw,
Poland}
\author{Piotr Fronczak}
\affiliation{Faculty of Physics and Center of Excellence Complex Systems
Research \\Warsaw University of Technology \\ Koszykowa 75, PL--00-662 Warsaw,
Poland}
\date{\today}

\begin{abstract}
We study a supremacy  distribution in evolving Barabasi-Albert networks. The
supremacy $s_i$   of a node $i$ is defined as a total number of all nodes that
are younger than $i$ and can be connected to it by a directed path. For a
network with a characteristic parameter $m=1,2,3,\dots$ the supremacy of an
individual node increases with the network age as $t^{(1+m)/2}$ in an
appropriate scaling region.  It follows that there is a relation $s(k) \sim
k^{m+1}$ between a node degree $k$ and its supremacy $s$ and the supremacy
distribution $P(s)$ scales as $s^{-1-2/(1+m)}$. Analytic calculations basing on
a continuum theory of supremacy evolution and on a corresponding rate equation
have been confirmed by numerical simulations.
\end{abstract}

\pacs{89.75.-k, 02.50.-r} \keywords{evolving networks, scaling}
\maketitle

\section{Introduction}
During the last few years there has been a large interest in modeling of
networks \cite{BAa,BAb,0a,0b,Oc} and  several parameters describing the network
structure have been considered. The examples are: degree distribution $P(k)$
\cite{BAb,Ka}, mean path length \cite{7a,8,9a,Agata2}, betweenness centrality
(load) \cite{9a,b} or first and higher order clustering coefficients
\cite{c1,ffh1,Agata1}. Universal scaling has been observed for some of these
parameters in computer simulations and in real data describing such objects as
the Internet, WWW, scientific collaboration networks or food webs
\cite{0a,0b,Oc}. Here we introduce a new parameter that can play an important
role for description of a class of directed networks. We name the parameter a
{\it supremacy} since it describes the number of nodes that are subordinated to
a certain node. In the next Section we define our parameter and show its
relevance for different problems of complex networks, Sec. \ref{Cont} includes a
continuum theory for the supremacy time evolution $s_i(t)$ and the supremacy
probability distribution $P(s)$  in the Barabasi-Albert (BA) model with $m=1$,
in Sec. \ref{Rate} we find and solve  a corresponding rate equation while in
Sec. \ref{Scaling} a generalization of our problem for the BA model with $m>1$
is presented.

\section{The model}
\begin{figure}\epsfxsize=5cm\epsfbox{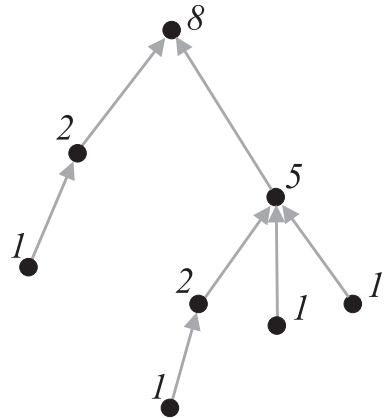}
\caption{Schematic illustration of the supremacy effects in
tree-like BA network with $m=1$. Numbers situated in the vicinity
of the nodes represent their supremacies.} \label{m1tree}
\end{figure}

Let us consider the BA network with the characteristic parameter $m=1$
\cite{BAa,BAb}. At the moment $t_i$ a node $i$ is created and it attaches to
some older node in the network according to the preferential attachment rule
(PAR). Then in the next time steps other nodes are created and are attached to
the node $i$ or to other nodes of the network following PAR. As a result at the
moment $t>t_i$ there is a subgraph in a form of tree $T(i,t)$ beginning in the
node $i$ and containing all nodes that are younger than the node $i$ and that
are connected to $i$ by directed paths as in Fig. \ref{m1tree}. If we assume
that the node $i$ represents a scientist  who wrote an important paper or a
politician who created an influential party \cite{jsp96,PhysA} we can consider
all nodes belonging to the tree as his/her successors. If the tree $T(i,t)$
contains $s_i$ nodes then the number $s_i$ is the measure of the supremacy or
the predominance of the node $i$ at time $t$. The  subgraph $T(i)$ can be also
interpreted  as a cluster of connected sites in the directed percolation problem
\cite{grass1,Cohen1,Cohen2,Agata3,n,d} and the supremacy of a node $i$ is just
the size of such a  cluster starting from the site $i$. Since the evolution of
the network is governed by PAR and all properties of the network are described
by some probability distributions we are interested in the supremacy
distribution $P(s)$ in the network.

\section{\label{Cont}Continuum theory of supremacy evolution and distribution for $m=1$}

To find the supremacy distribution $P(s)$ we follow the method that was
introduced in \cite{BAb} for calculation of degree distribution $P(k)$ in
evolving networks. We start from determining the time dependence of $s_i(t)$
assuming that it is a continuous real variable.  The supremacy of the node $i$
increases in time because new nodes can be  attached to any node of the tree
$T(i,t)$. Let nodes belonging to the tree $T(i,t)$ possess degrees $k_i^{(1)},
k_i^{(2)}, \ldots k_i^{(s_i)}$.

\begin{figure}\epsfxsize=8.5cm\epsfbox{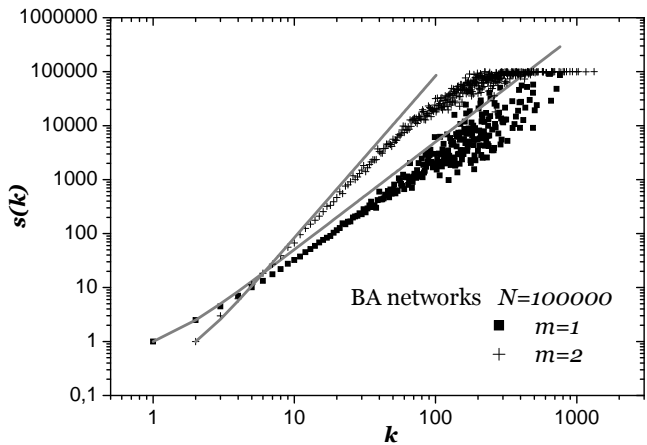}
\caption{Supremacy as a function the node degree. Solid lines represent
analytical predictions of $s(k)$ given by Eq.\ref{eq5} in the case of $m=1$ and
Eq.\ref{af1} in the case of $m=2$.} \label{sk}
\end{figure}

Using the PAR we can write the following equation for changes of
$s_i(t)$
\be
\frac{\partial s_i(t)}{\partial
t}=\sum_{l=1}^{s_i}\frac{k_i^{(l)}}{2t}=\frac{K(i)}{2t}\lab{eq1},
\end{equation}
where $K(i)=\sum_{l=1}^{s_i}k_i^{(l)}$ and we used the fact that
at the moment $t$ the sum of all nodes degrees in the whole
network equals to $2t$. On the other hand taking into account the
tree structure of the considered subgraph we can write the
supremacy $s_i$  as \be
s_i=1+\sum_{l=1}^{s_i}(k_i^{(l)}-1)=1+K(i)-s_i,\lab{eq2}\end{equation}
thus $ K(i)= 2 s_i-1$ and we have a simple equation \be
\frac{\partial s_i(t)}{\partial t}=\frac{2s_i-1}{2t},
\lab{eq3}\end{equation} with the solution \be
s_i(t)=\frac{1}{2}\left(\frac{t}{t_i}+1\right),\lab{eq4}\end{equation}
where we took into account the initial condition $s_i(t=t_i)=1$.
The solution (\ref{eq4}) means that the node supremacy increases
linearly in time comparing to the square root dependence of the
node degree \cite{BAb}, i.e. $k_i(t)=\sqrt{\frac{t}{t_i}} $.
Combining the last two results we get a simple relation between
the node supremacy and the node degree \be
s(k)=\frac{1}{2}\left[k^2+1\right] \lab{eq5}.\end{equation} In the
region $k\leq 100$  this formula fits  well to numerical
simulations presented in Fig.\ref{sk} while for larger $k$
differences between the analytic theory and the numerical
simulations are observed.

\begin{figure}\epsfxsize=8.5cm\epsfbox{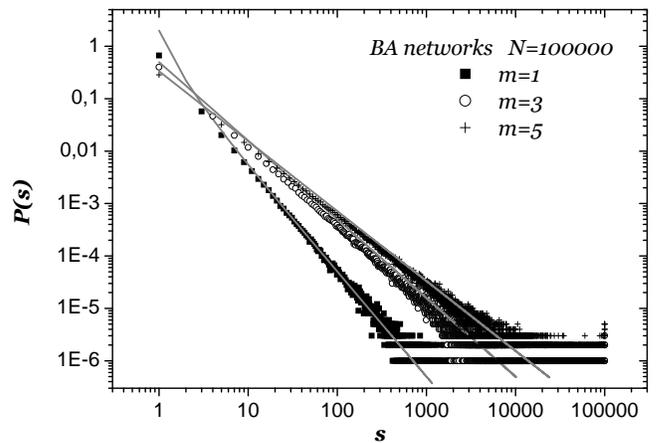}
\caption{Supremacy distribution in BA model. Solid lines represent analytical
predictions of $P(s)$ given by Eq.\ref{eq7} in the case of $m=1$ and
Eq.\ref{eq13} in the case of $m=3,$ $5$.} \label{ps}
\end{figure}

The probability density $P(s)$ for the supremacy distribution in the network
follows from the relation \be P(s_i)ds_i=\tilde{P}(t_i)dt_i,\end{equation} where
$\tilde{P}(t_i)=1/t$ is the distribution of nodes attachment times $t_i$ for a
network of age $t$. After a simple algebra we get \be
P(s_i)=\frac{1}{t}\left|\frac{\partial s_i}{\partial
t_i}\right|^{-1}=\frac{2}{(2s_i-1)^2} \label{eq7}.\end{equation} One can see
that the supremacy distribution  is a time independent function.  Fig.\ref{ps}
shows the comparison of the last equation to numerical data. Let us stress that
for $s\gg1$ the supremacy distribution scales as $P(s) \sim s^{-2}$ while the
degree distribution for BA model \cite{BAa,BAb} scales as $P(k)\sim k^{-3}$.

\section{\label{Rate}Rate-equation for supremacy distribution for $m=1$}
Now we show how to get the supremacy distribution using the rate-equation
approach that was introduced by Krapivsky, Redner and Leyvraz \cite{Ka} to study
networks degree distribution $P(k)$. Let $N(s,t)$ is the number of nodes
possessing the supremacy $s$ at time $t$. The rate equation for $N(s,t)$ is
\begin{eqnarray} \frac{\partial N(s,t)}{\partial
t}=\frac{[2(s-1)-1]N(s-1,t)-(2s-1)N(s,t)}{2t}\nonumber\\+\delta_{s,1}. \lab{eq8}
\end{eqnarray} The first term on the right-hand side of \eqn{eq8} corresponds to
creation of a new node with the supremacy $s$. The process is
proportional to the number of nodes with the supremacy $s-1$ and
the corresponding transition probability that follows from the PAR
and \eqn{eq2}. The second term corresponds to creation of a node
with a supremacy $s+1$, i.e. to destruction of a node with a
supremacy $s$ while the last term describes creation of a node
with a supremacy $s=1$. Writing $N(s,t)=P(s)N_0$ where $N_0=t$
corresponds to the total number of nodes at time $t$ and $P(s)$ is
the probability of a node with the supremacy value $s$ we get the
recursive equation \be P(s)=\frac{2s-3}{2s+1}P(s-1)
\mbox{\hspace{0.5cm} for }s\geq2 \lab{eq9},\end{equation} where
$P(1)=2/3$. The solution of \eqn{eq9} is \be
P(s)=\frac{2}{(2s-1)(2s+1)}. \lab{eq10}\end{equation} Note that
for $s\gg1$ the solution (\ref{eq10}) coincides   with the
solution (\ref{eq7}) that has been received in the limit of the
continuum theory.

\section{\label{Scaling}Scaling of supremacy distribution  for  $m>1$}
The peculiar feature of the BA model is the independence of the scaling exponent
characterizing the degree distribution $P(k)\sim k^{-3}$ from the model
parameter $m$ describing the number of links that are created by every new node.
Below we show that the scaling exponent of supremacy distribution depends on the
parameter $m$. If we neglect all loops existing in the BA network with the
characteristic parameter $m>1$ then  we can easy repeat our considerations from
Sec.\ref{Cont} and \ref{Rate}. Instead of \eqn{eq2} we get \be
s_i=1+\sum_{l=1}^{s_i}(k_i^{(l)}-m)=K(i)+1-ms_i,\lab{eq11}\end{equation} and
time evolution of the supremacy is described by \be
s_i(t)=\frac{m}{m+1}\left(\frac{t}{t_i}\right)^{\frac{m+1}{2}}+\frac{1}{m+1},\lab{eq12}\end{equation}
thus the relation between the degree and the supremacy is
\begin{equation}\label{af1}
s(k)=\frac{m}{m+1}\left(\frac{k}{m}\right)^{m+1}+\frac{1}{m+1}.
\end{equation}
It follows that for dense networks with $m \gg1$ the supremacy $s_i(t)$
increases in time much faster than the node degree $k_i(t)$. Fig. \ref{sk} shows
a comparison of the result  (\ref{af1}) to numerical data for $m=2$. One can see
that the predicted scaling of $s(k)$ breaks down completely for large values of
$k$ where the plot $s(k)$ saturate. The reason is the presence of loops that for
$m>1$ appear in the network and that have been neglected in our approach. If
$m>1$  the result  (\ref{af1}) is valid mainly for vertices with a small degree
$k_i$ (and a small supremacy $s_i$) since  loops are sparse in small  clusters
starting from such nodes.  The saturation effect does not appear for the BA
model with the parameter $m=1$ where loops are absent.
\begin{figure}\epsfxsize=5.5cm\epsfbox{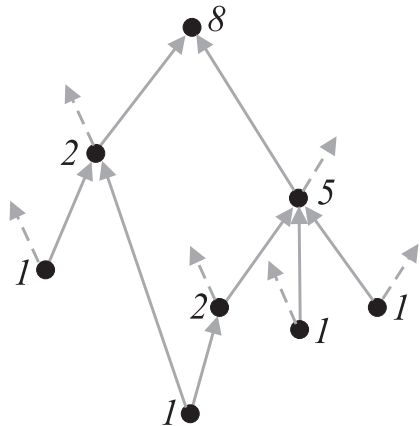}
\caption{Schematic illustration of  supremacy effects in BA network with $m=2$.
Solid arrows represent connections within the supremacy area / cluster of the
top vertex, whereas dashed arrows express connections pointing outside the
cluster. Note that there is a single loop in the cluster.}\label{m2tree}
\end{figure}

Taking into account (\ref{eq12}) we get the supremacy distribution
in the form
\begin{equation}
P(s_i)=\frac{1}{t}\left|\frac{\partial s_i} {\partial
t_i}\right|^{-1} =\frac{2}{m}\left[\frac{(m+1)s_i-1}{m}
\right]^{-\frac{m+3}{m+1}}\lab{eq13}.
\end{equation}
We see that the scaling exponent for the supremacy distribution equals to
$\delta=-1-2/(1+m)$ and in contrast to the scaling exponent of degree
distribution it depends on the parameter $m$. The result (\ref{eq13}) is in a
good agreement with numerical simulation for BA networks, see Fig. \ref{ps}.
\par The rate-equation for $m>1$
is  similar to \eqn{eq8}, i.e. \begin{eqnarray} \frac{\partial N(s,t)}{\partial
t}=\frac{[(1+m)(s-1)-1]N(s-1,t)}{2t}\nonumber\\-\frac{[(1+m)s-1)]N(s,t)}{2t}+\delta_{s,1}
\lab{eq14}\end{eqnarray}   The resulting solution for the probability $P(s)$ can
be written as  the following product \be
P(s)=\frac{2}{m+2}\prod_{i=2}^s\frac{\left[(i-1)(m+1)-1\right]}{\left[i(m+1)+1\right]}
\label{prod}\end{equation} for $s>1$ where $P(1)=2/(m+2)$. For dense networks
$m\gg1$ the solution (\ref{prod}) can be approximately written as \be
P(s)\simeq\frac{2}{ms}
\end{equation} what coincides with (\ref{eq13}).

\section{Conclusions}
In conclusion, we introduced a universal parameter (a supremacy) that describes
vertices in directed networks. The parameter equals to the size of a cluster
starting from the site in a directed percolation model.  We have shown that for
Barabasi-Albert model there is a relation between the supremacy and the vertex
degree. It follows that there are universal scaling laws describing the time
evolution of the supremacy and corresponding supremacy distributions in BA
models. On the contrary to the scaling results for nodes degree the
corresponding scaling exponents of supremacy depend on the characteristic model
parameter $m$. Numerical simulations are in good agreement with analytical
estimations for node with  a small and medium supremacy especially for the case
$m=1$ where no loops are present in the system.

\section{Acknowledgments}
The paper was supported by the special program of the Warsaw University of
Technology {\it Dynamics of Complex Systems}.

\end{document}